%Paper: hep-th/9305031
%From: BASSETTO@padova.infn.it
%Date: Sat, 8 May 1993 11:42:50 +0200 (WET-DST)

%
%---------------------------------------------------------------------
%--------------------- automacro, versione 1.1 -----------------------
%-------------------------  10 / 2 / 89  -----------------------------
%---------------------------------------------------------------------
%
\catcode`@=11
%
%------------------------- comandi riservati ---------------------------
%
\def\b@lank{ }

\newif\if@simboli
\newif\if@riferimenti
\newif\if@bozze

\newwrite\file@simboli
\def\simboli{
    \immediate\write16{ !!! Genera il file \jobname.SMB }
    \@simbolitrue\immediate\openout\file@simboli=\jobname.smb}

\newwrite\file@ausiliario
\def\riferimentifuturi{
    \immediate\write16{ !!! Genera il file \jobname.AUX }
    \@riferimentitrue\openin1 \jobname.aux
    \ifeof1\relax\else\closein1\relax\input\jobname.aux\fi
    \immediate\openout\file@ausiliario=\jobname.aux}

\def\bozze{\@bozzetrue}

\newcount\eq@num\global\eq@num=0
\newcount\sect@num\global\sect@num=0

\newif\if@ndoppia
\def\numerazionedoppia{\@ndoppiatrue\gdef\la@sezionecorrente{\the\sect@num}}

\def\se@indefinito#1{\expandafter\ifx\csname#1\endcsname\relax}
\def\spo@glia#1>{} % si applica a \meaning\xxxxx; butta via tutto quello
                   % che produce \meaning fino al carattere >
                   % (v. manuale TeX, pag. 382, \strip#1>{}).

\newif\if@primasezione
\@primasezionetrue

\def\s@ection#1\par{\immediate
    \write16{#1}\if@primasezione\global\@primasezionefalse\else\goodbreak
    \vskip\spaziosoprasez\fi\noindent
    {\bf#1}\nobreak\vskip\spaziosottosez\nobreak\noindent}
%
%------------------------------ a disp. dell'utente:  sezioni -------------

\def\sezpreset#1{\global\sect@num=#1
    \immediate\write16{ !!! sez-preset = #1 }   }

\def\spaziosoprasez{50pt plus 60pt}
\def\spaziosottosez{15pt}

\def\sref#1{\se@indefinito{@s@#1}\immediate\write16{ ??? \string\sref{#1}
    non definita !!!}
    \expandafter\xdef\csname @s@#1\endcsname{??}\fi\csname @s@#1\endcsname}

\def\autosez#1#2\par{
    \global\advance\sect@num by 1\if@ndoppia\global\eq@num=0\fi
    \xdef\la@sezionecorrente{\the\sect@num}
    \def\usa@getta{1}\se@indefinito{@s@#1}\def\usa@getta{2}\fi
    \expandafter\ifx\csname @s@#1\endcsname\la@sezionecorrente\def
    \usa@getta{2}\fi
    \ifodd\usa@getta\immediate\write16
      { ??? possibili riferimenti errati a \string\sref{#1} !!!}\fi
    \expandafter\xdef\csname @s@#1\endcsname{\la@sezionecorrente}
    \immediate\write16{\la@sezionecorrente. #2}
    \if@simboli
      \immediate\write\file@simboli{ }\immediate\write\file@simboli{ }
      \immediate\write\file@simboli{  Sezione
                                  \la@sezionecorrente :   sref.   #1}
      \immediate\write\file@simboli{ } \fi
    \if@riferimenti
      \immediate\write\file@ausiliario{\string\expandafter\string\edef
      \string\csname\b@lank @s@#1\string\endcsname{\la@sezionecorrente}}\fi
    \goodbreak\vskip 48pt plus 60pt
    \noindent\if@bozze\llap{\tt#1\quad }\fi
      {\bf\the\sect@num.\quad #2}\par\nobreak\vskip 15pt
    \nobreak\noindent}

\def\semiautosez#1#2\par{
    \gdef\la@sezionecorrente{#1}\if@ndoppia\global\eq@num=0\fi
    \if@simboli
      \immediate\write\file@simboli{ }\immediate\write\file@simboli{ }
      \immediate\write\file@simboli{  Sezione ** : sref.
          \expandafter\spo@glia\meaning\la@sezionecorrente}
      \immediate\write\file@simboli{ }\fi
    \s@ection#2\par}

%------------------------------ a disp. dell'utente:  equazioni -----------

\def\eqpreset#1{\global\eq@num=#1
     \immediate\write16{ !!! eq-preset = #1 }     }

\def\eqref#1{\se@indefinito{@eq@#1}
    \immediate\write16{ ??? \string\eqref{#1} non definita !!!}
    \expandafter\xdef\csname @eq@#1\endcsname{??}
    \fi\csname @eq@#1\endcsname}

\def\eqlabel#1{\global\advance\eq@num by 1
    \if@ndoppia\xdef\il@numero{\la@sezionecorrente.\the\eq@num}
       \else\xdef\il@numero{\the\eq@num}\fi
    \def\usa@getta{1}\se@indefinito{@eq@#1}\def\usa@getta{2}\fi
    \expandafter\ifx\csname @eq@#1\endcsname\il@numero\def\usa@getta{2}\fi
    \ifodd\usa@getta\immediate\write16
       { ??? possibili riferimenti errati a \string\eqref{#1} !!!}\fi
    \expandafter\xdef\csname @eq@#1\endcsname{\il@numero}
    \if@ndoppia
       \def\usa@getta{\expandafter\spo@glia\meaning
       \la@sezionecorrente.\the\eq@num}
       \else\def\usa@getta{\the\eq@num}\fi
    \if@simboli
       \immediate\write\file@simboli{  Equazione
            \usa@getta :  eqref.   #1}\fi
    \if@riferimenti
       \immediate\write\file@ausiliario{\string\expandafter\string\edef
       \string\csname\b@lank @eq@#1\string\endcsname{\usa@getta}}\fi}

\def\autoreqno#1{\eqlabel{#1}\eqno(\csname @eq@#1\endcsname)
       \if@bozze\rlap{\tt\quad #1}\fi}
\def\autoleqno#1{\eqlabel{#1}\leqno\if@bozze\llap{\tt#1\quad}
       \fi(\csname @eq@#1\endcsname)}
\def\eqrefp#1{(\eqref{#1})}
\def\numeriadestra{\let\autoeqno=\autoreqno}
\def\numeriasinistra{\let\autoeqno=\autoleqno}
\numeriadestra
%--------------- bibliografia automatica: riservati ----------------------

\newcount\cit@num\global\cit@num=0

\newwrite\file@bibliografia
\newif\if@bibliografia
\@bibliografiafalse

\def\lp@cite{[}
\def\rp@cite{]}
\def\trap@cite#1{\lp@cite #1\rp@cite}
\def\lp@bibl{[}
\def\rp@bibl{]}
\def\trap@bibl#1{\lp@bibl #1\rp@bibl}

\def\refe@renza#1{\if@bibliografia\immediate        % scrive su .BIB
    \write\file@bibliografia{
    \string\item{\trap@bibl{\cref{#1}}}\string
    \bibl@ref{#1}\string\bibl@skip}\fi}

\def\ref@ridefinita#1{\if@bibliografia\immediate\write\file@bibliografia{
    \string\item{?? \trap@bibl{\cref{#1}}} ??? tentativo di ridefinire la
      citazione #1 !!! \string\bibl@skip}\fi}

\def\bibl@ref#1{\se@indefinito{@ref@#1}\immediate
    \write16{ ??? biblitem #1 indefinito !!!}\expandafter\xdef
    \csname @ref@#1\endcsname{ ??}\fi\csname @ref@#1\endcsname}

\def\c@label#1{\global\advance\cit@num by 1\xdef            % assegna il numero
   \la@citazione{\the\cit@num}\expandafter
   \xdef\csname @c@#1\endcsname{\la@citazione}}

\def\bibl@skip{\vskip 0truept}

%------------------------ bibl. automatica: a disp. dell'utente ------------

\def\stileincite#1#2{\global\def\lp@cite{#1}\global
    \def\rp@cite{#2}}
\def\stileinbibl#1#2{\global\def\lp@bibl{#1}\global
    \def\rp@bibl{#2}}

\def\citpreset#1{\global\cit@num=#1
    \immediate\write16{ !!! cit-preset = #1 }    }

\def\autobibliografia{\global\@bibliografiatrue\immediate
    \write16{ !!! Genera il file \jobname.BIB}\immediate
    \openout\file@bibliografia=\jobname.bib}

\def\cref#1{\se@indefinito                  % se indefinito definisce
   {@c@#1}\c@label{#1}\refe@renza{#1}\fi\csname @c@#1\endcsname}

\def\cite#1{\trap@cite{\cref{#1}}}                  %  [5]
\def\ccite#1#2{\trap@cite{\cref{#1},\cref{#2}}}     %  [5,6]
\def\ncite#1#2{\trap@cite{\cref{#1}--\cref{#2}}}    %  [5-8] senza definire
\def\upcite#1{$^{\,\trap@cite{\cref{#1}}}$}               % ^[5]
\def\upccite#1#2{$^{\,\trap@cite{\cref{#1},\cref{#2}}}$}  % ^[5,6]
\def\upncite#1#2{$^{\,\trap@cite{\cref{#1}-\cref{#2}}}$}  % ^[5-8] senza def.

\def\clabel#1{\se@indefinito{@c@#1}\c@label           % sola definizione
    {#1}\refe@renza{#1}\else\c@label{#1}\ref@ridefinita{#1}\fi}
                     % def. doppia
       % def. tripla

\def\biblskip#1{\def\bibl@skip{\vskip #1}}           % spaziatura nella bibl.

\def\insertbibliografia{\if@bibliografia             % scrive la bibliografia
    \immediate\write\file@bibliografia{ }
    \immediate\closeout\file@bibliografia
    \catcode`@=11\input\jobname.bib\catcode`@=12\fi}

%--------- per comporre il file con la bibliografia --------------

\def\commento#1{\relax}
\def\biblitem#1#2\par{\expandafter\xdef\csname @ref@#1\endcsname{#2}}

% ricordare: una lista in chiaro della bibliografia si
% ottiene eseguendo $ TEX BIBLIST

%------------------------------ F I N E ---------------------------------
\catcode`@=12

\catcode`@=11
%
%---------------------------- \lsim \gsim ----------------------------------
%
%    Simboli di minore o circa uguale, maggiore o circa uguale.
%
\def\lsim{\mathchoice
  {\mathrel{\lower.8ex\hbox{$\displaystyle\buildrel<\over\sim$}}}
  {\mathrel{\lower.8ex\hbox{$\textstyle\buildrel<\over\sim$}}}
  {\mathrel{\lower.8ex\hbox{$\scriptstyle\buildrel<\over\sim$}}}
  {\mathrel{\lower.8ex\hbox{$\scriptscriptstyle\buildrel<\over\sim$}}} }
\def\gsim{\mathchoice
  {\mathrel{\lower.8ex\hbox{$\displaystyle\buildrel>\over\sim$}}}
  {\mathrel{\lower.8ex\hbox{$\textstyle\buildrel>\over\sim$}}}
  {\mathrel{\lower.8ex\hbox{$\scriptstyle\buildrel>\over\sim$}}}
  {\mathrel{\lower.8ex\hbox{$\scriptscriptstyle\buildrel>\over\sim$}}} }
\def\croce{\displaystyle / \kern-0.2truecm\hbox{$\backslash$}}
\def\lqua{\lower4pt\hbox{\kern5pt\hbox{$\sim$}}\raise1pt
\hbox{\kern-8pt\hbox{$<$}}~}
\def\gqua{\lower4pt\hbox{\kern5pt\hbox{$\sim$}}\raise1pt
\hbox{\kern-8pt\hbox{$>$}}~}
\def\mma{\lower1pt\hbox{\kern5pt\hbox{$\scriptstyle <$}}\raise2pt
\hbox{\kern-7pt\hbox{$\scriptstyle >$}}~}
\def\mmb{\lower1pt\hbox{\kern5pt\hbox{$\scriptstyle >$}}\raise2pt
\hbox{\kern-7pt\hbox{$\scriptstyle <$}}~}
\def\mmc{\lower4pt\hbox{\kern5pt\hbox{$<$}}\raise1pt
\hbox{\kern-8pt\hbox{$>$}}~}
\def\mmd{\lower4pt\hbox{\kern5pt\hbox{$>$}}\raise1pt
\hbox{\kern-8pt\hbox{$<$}}~}
\def\croce{\displaystyle / \kern-0.2truecm\hbox{$\backslash$}}
%
%
%---------------------------  \quadratello ---------------------------------
%
%
\def\quad@rato#1#2{{\vcenter{\vbox{
        \hrule height#2pt
        \hbox{\vrule width#2pt height#1pt \kern#1pt \vrule width#2pt}
        \hrule height#2pt} }}}
\def\quadratello{\mathchoice
\quad@rato5{.5}\quad@rato5{.5}\quad@rato{3.5}{.35}\quad@rato{2.5}{.25} }
%
%------------------------ caratteri grassetto speciali -----------
%
\font\s@=cmss10\font\s@b=cmbx8
\def\reali{{\hbox{\s@ l\kern-.5mm R}}}
\def\m{{\hbox{\s@ l\kern-.5mm M}}}
\def\k{{\hbox{\s@ l\kern-.5mm K}}}
\def\naturali{{\hbox{\s@ l\kern-.5mm N}}}
\def\interi{{\mathchoice
 {\hbox{\s@ Z\kern-1.5mm Z}}
 {\hbox{\s@ Z\kern-1.5mm Z}}
 {\hbox{{\s@b Z\kern-1.2mm Z}}}
 {\hbox{{\s@b Z\kern-1.2mm Z}}}  }}
\def\complessi{{\hbox{\s@ C\kern-1.7mm\raise.4mm\hbox{\s@b l}\kern.8mm}}}
\def\toro{{\hbox{\s@ T\kern-1.9mm T}}}
\def\unity{{\hbox{\s@ 1\kern-.8mm l}}}
%
%------------------------- bold math. it. --------------------------
%
\font\bold@mit=cmmib10
\def\setbmit{\textfont1=\bold@mit}
\def\bmit#1{\hbox{\textfont1=\bold@mit$#1$}}
%
%-----------------------------------------------------------------
\catcode`@=12

\magnification=\magstep1
\hsize 15.5truecm
\vsize 23truecm
\baselineskip 16pt
\parindent=0.5truecm
\overfullrule=0pt
\numerazionedoppia
\autobibliografia
\stileincite()
\stileinbibl{})
\def\spaziosoprasez{10pt plus 5pt}
\def\spaziosottosez{5pt}

\nopagenumbers
\footline{\ifnum\pageno>1 \hss\tenrm\folio\hss \else\hfil\fi}
\pageno=1

\def\kbar{\hbox{$k$}\kern-0.2truecm\hbox{$/$}}
\def\nbar{\hbox{$n$}\kern-0.23truecm\hbox{$/$}}
\def\pbar{\hbox{$p$}\kern-0.18truecm\hbox{$/$}}
\def\parz{\hbox{\hbox{${\partial}$}}\kern-1.7mm{\hbox{${/}$}}}
\def\A{\hbox{\hbox{${A}$}}\kern-1.8mm{\hbox{${/}$}}}

\null
\centerline{\bf Non Perturbative Solutions and Scaling Properties of
Vector,}
\centerline{\bf Axial--Vector Electrodynamics in $1+1$ Dimensions}
\vskip 1truecm
\centerline{by}
\vskip 0.5truecm
\centerline{A. Bassetto}
\vskip 0.3truecm
\centerline{\it Dipartimento di Fisica ``G. Galilei", Via Marzolo 8 -
35131 Padova, Italy}

\centerline{\it INFN, Sezione di Padova, Italy}
\vskip 0.5truecm
\centerline{L. Griguolo}
\vskip 0.3truecm
\centerline{\it SISSA, via Beirut 2-34100 Trieste, Italy}

\centerline{\it INFN, Sezione di Trieste, Italy}
\vskip 0.5 truecm
\centerline{P. Zanca}
\vskip 0.3 truecm
\centerline{\it Dipartimento di Fisica ``G. Galilei", Via Marzolo 8 -
35131 Padova, Italy}
\vskip 1.0 truecm
\noindent
\underbar{Abstract:} We study by non perturbative techniques a vector,
axial--vector theory characterized by a parameter which interpolates
between pure vector and chiral Schwinger models. Main results are two
windows in the space of parameters which exhibit acceptable solutions. In
the first window we find a free massive and a free massless bosonic
excitations and interacting left--right fermions endowed with asymptotic
\hbox{states}, which feel  however a long range interaction.
In the second window
the massless bosonic excitation is a negative norm state which can be
consistently expunged from the ``physical" Hilbert space; fermions are
confined. An intriguing feature of our model occurs in the first
window where we find that fermionic correlators scale at both short and
long distances, but with different critical exponents.
The infrared limit in the fermionic
sector is nothing but a dynamically generated massless Thirring model.

\vfill
\eject

\autosez{int} \underbar{Introduction}

\par
Quantum field theories in 1--space, 1--time dimensions are intensively
studied in recent years owing to their peculiarity of being exactly
solvable both by functional and by operatorial techniques.
\noindent
{}From a practical point of view they find interesting applications in string
models, while behaving as useful theoretical laboratories in which many
features, present also in higher dimensional theories, can be directly
tested. In addition 2--dimensional models possess a quite peculiar infrared
structure on their own.

Historically the first 2--dimensional model was proposed by
Thirring \upcite{Thi58}, describing a pure fermionic current--current
interaction.
The interest suddenly increased 4 years later, when Schwinger
\upcite{Sch62} was able
to obtain an exact solution for 2--dimensional electrodynamics with
massless spinors. This models is so rich of interesting and
surprising features, like e. g. dynamical generation of a mass for the
vector field, fermion confinement,etc., that, after thirthy years, it
is still the subject of several investigations.

The chiral generalization of this model, first examined by Jackiw and
Rajaraman \upcite{Jac85}, allowed to draw very important conclusions concerning
theories with ``anomalies", i. e. the occurrence of symmetry breakings by
quantum effects.
They were able to show that, taking advantage of the arbitrariness in the
(non perturbative) regularization of the fermionic determinant, it was
possible to recover a unitary theory even in the presence of a gauge
anomaly.

The literature on the subject is so huge, that it is impossible to refer it
adequately; we just quote the book by Abdalla, Abdalla and
Rothe \upcite{Abd91}, where many references can be found.
\smallskip
In this paper we study in a two dimensional space with trivial topology
a family of theories which interpolate between vector and chiral
Schwinger models according to a parameter {\it r}, which tunes the ratio of the
axial to vector coupling.
\noindent
Our treatment will therefore depend on two parameters: {\it r} and {\it a},
{\it a} being the
constant involved in the regularization of the fermionic determinant.

In sect. 2 we obtain, by means of a functional approach, the
correlation functions for bosons, fermions and fermionic condensates.
We find two allowed windows for the parameters {\it r} and {\it a}.
The first window was also partially studied in a similar context in
\upccite{Hal86}{Miy88}. In this window
the bosonic sector consists of two ``physical" quanta, a free massive
and a free massless excitation. The fermionic sector is much more
interesting: both left and right spinors exhibit a propagator decreasing at
very large distances, indicating the presence of asymptotic states which
however feel the long range interaction mediated by the massless boson.

The solution interpolates between two conformal invariant theories at small
and large distances, respectively, with different critical exponents.
This very interesting feature of our model
is under investigation and the results will be
reported in a forthcoming paper.

For $r=0$ one recovers the vector Schwinger model; for $r=\pm 1$ one gets
the chiral model, where, in particular, one of the fermions is free.

The second window is characterized in the bosonic sector by a ``physical"
massive excitation and by a massless negative norm state (``ghost").
Both quanta are free; one can define a stable Hilbert space of states in
which the ``ghost" does not appear. However no asymptotic states for
fermions are available in this case; their correlation function increases
with distance, giving rise to a confinement phenomenon.

All those features are confirmed and further elucidated in subsequent
sections: in sect. 3 the bosonic sector is investigated by means of
operators which are canonically quantized according to a Dirac bracket
formalism \upcite{Dir50};
the structure of Hilbert space of states is discussed. Sect. 4
deals with the fermionic sector: fermionic operators are explicitly
constructed, quantized, and correlation functions are examined, also in
connection with the relevant equations of motion.
We also discuss their behaviour under symmetries and related charges.

In sect. 5 we show that the fermionic correlation functions of our model at
long distances exactly become the ones of a massless Thirring model, which
is the conformal invariant infrared limit of our theory. This deep relation
is present in the expression of operator fields and charges.

Sect. 6 contains final conclusions, while some technical details are
deferred to the Appendices.
\vfill

\autosez{the}\underbar{The path--integral formulation}

\par
The model, characterized by the classical Lagrangian
$$
\eqalign{{\cal L}=&-{1\over 4}F_{\mu\nu}F^{\mu\nu}+\bar\psi
i\parz\psi+e\bar\psi\gamma^\mu\psi A_\mu+\cr
&+re\bar\psi\gamma^\mu\gamma^5\psi A_\mu\cr}
\autoeqno{eq21}
$$
\noindent
will be quantized in this section following the path--integral method. In
\eqrefp{eq21} $F_{\mu\nu}$ is the usual field tensor, $A_\mu$ the vector
potential and $\psi$ a massless spinor. The quantity $r$ is a real
parameter interpolating between the vector $(r=0)$ and the chiral $(r=\pm
1)$ Schwinger models. Our notations are
$$
\eqalign{
&g_{00}=-g_{11}=1, \qquad\qquad \epsilon^{01}=-\epsilon_{01}=1,\cr
&\gamma^0=\sigma_1,\qquad \gamma^1=-i\sigma_2,\qquad
\gamma^5=\sigma_3,\qquad \tilde\partial_\mu=\epsilon_{\mu\nu}\partial^\nu,\cr}
\autoeqno{eq22}
$$
\noindent
$\sigma_i$ being the usual Pauli matrices.

The classical Lagrangian
\eqrefp{eq21} is invariant under the local trasformations
$$
\eqalign{
&\psi'(x)=\exp\left[ie\left(1+r\gamma^5\right)\Lambda(x)\right]\psi(x),\cr
&A'_\mu(x)=A_\mu(x)+\partial_\mu\Lambda.\cr}
\autoeqno{eq23}
$$
However, as is well known, it is impossible to make the fermionic
functional measure simultaneously invariant under  vector and axial vector
gauge transformations; as a consequence, for $r\not=0$ the quantum theory
will exhibit anomalies.

The Green function generating functional is
$$
W[J_\mu,\bar\eta, \eta]={\cal N}\int{\cal D}(A_\mu, \bar\psi,
\psi)e^{i\int({\cal L}+{\cal L}_s)d^2x},
\autoeqno{eq24}
$$
\noindent
where ${\cal N}$ is a normalization constant and
$$
{\cal L}_{s}=J_\mu A^\mu+\bar\eta\psi+\bar\psi\eta,
\autoeqno{eq25}
$$
\noindent
$J_\mu$, $\eta$ and $\bar\eta$ being vector and spinor sources
respectively.

The integration over the fermionic degrees of freedom can be
performed, leading to the expression
$$
\eqalign{
W[J_\mu, \eta, \bar\eta]=&{\cal N}\int{\cal D}(A_\mu,\phi)e^{i\int d^2x
{\cal L}_{eff}(A_\mu, \phi)}\cr
&e^{i\int d^2 xJ_\mu A^\mu} e^{-i\int
d^2xd^2y\bar\eta(x)S(x,y;A_\mu)\eta(y)}\cr}
\autoeqno{eq26}
$$
\noindent
where
$$
{\cal L}_{eff}=-{1\over 4}F_{\mu\nu}F^{\mu\nu}+{ae^2\over 2\pi}A_\mu
A^\mu+{1\over 2}\partial_\mu\phi\partial^\mu\phi+{e\over
\sqrt{\pi}}A^\mu(\tilde\partial_\mu-r\partial_\mu)\phi,
\autoeqno{eq27}
$$
\noindent
$\phi$ being a scalar field we have introduced in order to have a local
${\cal L}_{eff}$ and {\it a} the subtraction parameter reflecting the
well--known regularization ambiguity of the fermionic determinant
\upcite{Jac85}.

The quantity $ S(x,y;A_\mu)$ in \eqrefp{eq26} is the fermionic propagator
in the presence of the potential $A_\mu$, which will be computed later on
by using standard decoupling techniques.

For the moment we let the sources $\eta$ and $\bar\eta$ vanish and consider
the bosonic sector of the model for different values of the parameters $r$
and $a$. In this sector the effective Lagrangian is quadratic in the
fields; this means an essentially free (although non local) theory.

First functionally integrating over $\phi$ and then over $A_\mu$, we easily
obtain
$$
W[J_\mu, 0, 0]=\exp\left[-{1\over 2}\int d^2
xJ^\mu(K^{-1})_{\mu\nu}J^\nu\right],
\autoeqno{eq28}
$$
\noindent
where
$$
K_{\mu\nu}=g_{\mu\nu}\left(\quadratello+{e^2\over \pi}(1+a)\right)-\left(1+
{e^2\over \pi}{1+r^2\over \quadratello}\right)\partial_\mu\partial_\nu
+{e^2\over
\pi}{r\over
\quadratello}\left(\tilde\partial_\mu\partial_\nu
+\tilde\partial_\nu\partial_\mu\right)
\autoeqno{eq29}
$$
\noindent
and, consequently,
$$
\eqalign{
(K^{-1})_{\mu\nu}\equiv D_{\mu\nu} & ={1\over
\quadratello+m^2} \left[g_{\mu\nu}+{\quadratello+{e^2\over \pi}(1+r^2)\over
{e^2\over \pi} (a-r^2)} {\partial_\mu\partial_\nu\over
\quadratello}+\right.\cr
&\left. +{r\over r^2-a}{1\over
\quadratello} (\tilde\partial_\mu\partial_\nu+\tilde\partial_\nu\partial_\mu)
\right].\cr}
\autoeqno{eq210}
$$
We have introduced the quantity
$$
m^2={e^2\over \pi}{a\left(1+a-r^2\right)\over a-r^2} ,
\autoeqno{eq211}
$$
\noindent
which is to be interpreted as a dynamically generated mass in the
theory;
$D_{\mu\nu}$ has a pole there $\sim(k^2-m^2+i\epsilon)^{-1}$, with
causal prescription, as usual. We note that
$D_{\mu\nu}$ exhibits also a pole at $k^2=0$.

Eqs.
\eqrefp{eq210} and \eqrefp{eq211} generalize the well--known results of the
vector and chiral Schwinger models.
As a matter of fact, setting first $r=0$ and then $a=0$ we recover for
$m^2$ the value ${e^2\over \pi}$ of the (gauge invariant version of the)
vector Schwinger model.
The kinetic term $K_{\mu\nu}$ becomes a projection operator
$$
K_{\mu\nu}\left(a=0, r=0\right)=\left(\quadratello
+m^2\right)\left(g_{\mu\nu}-{\partial_\mu\partial_\nu\over
\quadratello}\right),
\autoeqno{eq212}
$$
\noindent
which can only be inverted after imposing a gauge fixing.
In other words the limit $r=0$, $a=0$ in \eqrefp{eq210} is singular, as it
should, as gauge invariance is indeed recovered.

When $r=\pm 1$, we obtain the two equivalent formulations of the chiral
Schwinger model; \eqrefp{eq211} becomes
$$
m^2={e^2\over \pi}{a^2 \over a-1}.
\autoeqno{eq213}
$$

To avoid tachyons, we must require $a>1$.
\noindent
Gauge invariance is definitely lost, and \eqrefp{eq210} becomes
$$
\eqalign{
D_{\mu\nu}&={1\over \quadratello+m^2}\left[g_{\mu\nu}+{1\over
a-1}\left({\pi\over e^2}+{2\over
\quadratello}\right)\partial_\mu\partial_\nu \mp\right.\cr
&\left.\mp{1\over
a-1}{\tilde\partial_\mu\partial_\nu+\tilde\partial_\nu\partial_\mu\over
\quadratello}\right].\cr}
\autoeqno{eq214}
$$

The limit $a\rightarrow 1$ is singular in \eqrefp{eq213}.
Nevertheless a definite expression can be obtained for the propagator
$$
\eqalign{
D_{\mu\nu}\mid_{a=1}&={\pi\over e^2}\left[\left({\pi\over e^2}+{2\over
\quadratello}\right)\partial_\mu\partial_\nu\mp{\tilde\partial_\mu
\partial_\nu+\tilde\partial_\nu\partial_\mu\over \quadratello}\right]=\cr
&={\pi\over
e^2}{\left(\partial_\mu+\tilde\partial_\mu\right)\left(\partial_\nu+
\tilde\partial_\nu\right)\over \quadratello},\cr}
\autoeqno{eq215}
$$
\noindent
where in the last equality ``contact terms" have been disregarded. They
correspond indeed to imposing different boundary conditions on the fields.

Going back to the general expression \eqrefp{eq211} we remark that the
condition $m^2>0$, which is necessary to avoid the presence of tachyons
in the theory, allows two windows:
$$
\eqalign{
1)\qquad\qquad\qquad & a>r^2,\cr
2)\qquad\qquad\qquad & 0<a<r^2-1 \qquad\qquad {\rm or} \qquad\qquad
r^2-1<a<0,\cr}
\autoeqno{eq216}
$$
\noindent
for the parameters $(a,r)$. Only the first window has been considered so
far in the literature, to our knowledge.

By taking in \eqrefp{eq210} the residue at the pole $k^2=m^2$, one gets
$$
Res~D_{\mu\nu}\mid_{k^2=m^2}={1\over m^2}T_{\mu\nu}(k),
\autoeqno{eq217}
$$
\noindent
$T_{\mu\nu}$ being a positive semidefinite degenerate quadratic form in the
parameters $(a,r)$. One eigenvalue vanishes, corresponding to a decoupling
of the would--be related excitation, the other is positive and can be
interpreted in both windows as the presence of a vector particle with
a rest mass given by the positive square root of \eqrefp{eq211} and
positive residue at the pole in agreement with the unitary condition.
\noindent
This state decouples in the limit $a=r^2$.
\noindent
There is also a massless degree of freedom with
$$
Res~D_{\mu\nu}\mid_{k^2=0}={\pi\over e^2
a(1+a-r^2)}\left[\left(1+r^2\right)k_\mu k_\nu-r\left(\tilde k_\mu
k_\nu+\tilde k_\nu k_\mu\right)\right]\mid_{k^2=0}.
\autoeqno{eq218}
$$

One can casily realize that again the quadratic form at the numerator is
positive semidefinite for any value of $r$. The poles at $k^2=m^2$ and
$k^2=0$ exhaust the singularities of $D_{\mu\nu}$.

Let us consider the situation in the two windows.
The first window does not deserve particular comments at this stage. No
ghost is present at $k^2=0$, as one eigenvalue of the residue matrix
vanishes and the other is positive, corresponding to a ``physical"
excitation.
The second window does entail no news concerning the state with mass $m$.
The situation is different however when considering the pole at $k^2=0$. We
have indeed a negative residue in this case corresponding to a ``ghost"
excitation (particle with a negative probability). The theory can be
accepted only if this excitation can be consistently excluded from a
positive norm Hilbert space of states, which is stable under time
evolution. This point will be reconsidered when we shall solve the model in
the framework of a canonical quantization.

To draw definite conclusions from this path--integral approach, it is worth
considering at this stage the fermionic sector.
The bosonic world is rather dull indeed, consisting only of free
excitations.

We go back to the general expression \eqrefp{eq26} in which fermionic
sources are on. We have now to consider the fermionic propagator in the
field $A_\mu$, which obeys the equation
$$
\left[i\parz+e\left(1-r\gamma^5\right)\A\right] S\left(x,
y;A_\mu\right)=\delta^2\left(x-y\right),
\autoeqno{eq219}
$$
\noindent
with causal boundary conditions.
\noindent
Let us also introduce the free propagator $S_0$
$$
i\parz S_0(x)=\delta^2(x)
\autoeqno{eq220}
$$
\noindent
with the solution
$$
S_0=\int{d^2 k\over (2\pi)^2}{\kbar\over k^2+i\epsilon} e ^{-ikx}={1\over
2\pi}{\gamma_\mu x^\mu\over x^2-i\epsilon}.
\autoeqno{eq221}
$$

If we remember that any vector in two dimensions can be written as a sum of
a gradient and a curl part
$$
A_\mu=\partial_\mu\alpha+\tilde\partial_\mu\beta,
\autoeqno{eq222}
$$
\noindent
the following change of variables in \eqrefp{eq24}
$$
\psi=\exp
\left[ie\left(\alpha+\gamma^5\beta+r\beta+r\alpha\gamma^5\right)\right]\chi
\autoeqno{eq223}
$$
\noindent
realizes the decoupling of the fermions, leading to the expression for the
``left" propagator ( see Appendix A)
$$
\eqalign{
S^L(x-y)&\equiv\int{\cal D}\left(A_\mu,\phi\right) S^L\left(x,y;
A_\mu\right) e^{i\int d^2z{\cal L}_{eff}(A_\mu,\phi)}=\cr
&=S_0^L(x-y)Z_L \exp\left\{-{1\over 4}{(1-r^2)^2\over
a(a+1-r^2)}\ln\left[\tilde m^2\left(-(x-y)^2+i\epsilon\right)\right]-\right.\cr
&\left.-i\pi{a+1-r^2\over a(a-r^2)}\left(r-{a\over a+1-r^2}\right)^2
D\left(x-y,m\right)\right\},\cr}
\autoeqno{eq224}
$$
\noindent
where $\tilde m={me^\gamma\over 2}$, $D$ is the scalar Feynman
propagator: $D\equiv D_0$, with
$$
\eqalign{
D_{1-\omega}\left(x, m\right)&=- (\lambda^2)^{1-\omega}\int{d^{2\omega}
k\over (2\pi)^{2\omega}}{e^{-ikx}\over
k^2-m^2+i\epsilon}\cr
&={2i\over (4\pi)^{\omega}}
\left({\lambda^2 \sqrt {-x^2}\over 2m}\right)^{1-\omega}
K_{1-\omega}\left(m\sqrt{-x^2+i\epsilon}\right),\cr}
\autoeqno{eq225}
$$
\noindent
$\gamma$ being the Euler--Mascheroni constant.
For further developments it is useful to consider $2\omega$ dimensions
and to introduce a  mass parameter
$\lambda$ to balance dimensions. $Z_L$ is a (dimensionally
regularized) ultraviolet renormalization constant for the fermion wave
function
$$
Z_L=\exp\left[i{\pi(r-1)^2\over a-r^2} D_{1-\omega}(0,m)\right].
\autoeqno{eq226}
$$

The ``right" propagator can be obtained from \eqrefp{eq224} simply by
replacing $S_0^L$ with $S_0^R$ and changing the sign of the
parameter $r$.

First of all, we notice that for $r=1$ the ``left" fermion is free. The
same  happens to the ``right" fermion when $r=-1$.
Moreover we notice from \eqrefp{eq224} that the long range interaction
completely decouples for $r^2=1$.
As a consequence the interacting fermion (for instance the ``right" one for
$r=1$) asymptotically behaves like a free particle.

In general, at small values of $x^2$, the propagator $S^L$ has the
following behaviour
$$
S^L\sim_{x^2\rightarrow 0}C_0 x^+\left(-x^2+i\epsilon\right)^{-1-A}
\autoeqno{eq227}
$$
\noindent
with
$$
A={1\over 4}{(1-r)^2\over a-r^2}
\autoeqno{eq228}
$$
\noindent
and $C_0$ a suitable constant.

\noindent
We remark that the ultraviolet behaviour of the left fermion propagator
can be directly obtained from the ultraviolet renormalization constant
$$
\gamma_{\psi_{L}}=\lim_{\omega\rightarrow 1}{1\over 2}
\left(\lambda{\partial\over \partial\lambda}lnZ_L\right)=-{(1-r)^2\over
4(a-r^2)}
\autoeqno{eq229}
$$
and, of course, it coincides
with the one of the explicit solution \eqrefp{eq224}.

For large values of $x^2$ we get instead
$$
S^L\sim_{x^2\rightarrow-\infty}~C_\infty
x^+\left(-x^2+i\epsilon\right)^{-1-B}
\autoeqno{eq230}
$$
\noindent
where
$$
B={1\over 4}{(1-r^2)^2\over a(a+1-r^2)}
\autoeqno{eq231}
$$
\noindent
and $C_\infty$ another constant.

We shall see in sect.5 that \eqrefp{eq230} exactly coincides with the
fermionic propagator of the massless Thirring model.

In the first window $(a>r^2)$, both $A$ and $B$ are positive. The
propagator decreases at infinity indicating the possible existence
of asymptotic
states for fermions, which however feel the long range interaction mediated
by the massless excitation which is present in the bosonic spectrum.
The situation in the second window is much more intriguing. Here both $A$
and $B$ are negative. Moreover
$$
1+B={(2a+1-r^2)^2\over 4a(a+1-r^2)}<0
\autoeqno{eq232}
$$
leading to a propagator which increases when $x^2\rightarrow -\infty$.
We interprete this phenomenon as a sign of confinement. We recall indeed
that gauge invariance is broken and therefore the fermion propagator is
endowed of a direct physical meaning.
The unphysical massless bosonic excitation, which occurs in this window,
produces an anti--screening effect of a long range type.
Nevertheless no asymptotic freedom is expected $(A\not=0)$.

All this analysis will be confirmed and reinterpreted in a deeper way
when following a canonical procedure.

Propagators are not suitable to discuss the limiting case $a=r=0$
(vector Schwinger model) in which gauge invariance is restored.
There is however another interesting quantity which can be easily discussed
in a path--integral approach.
Let us introduce the scalar fermion condensate
$$
S(x)=N\left[\bar\psi(x)\psi(x)\right]
\autoeqno{eq233}
$$
\noindent
where $N$ means the finite part, after divergences have been
(dimensionally) regularized and renormalized.
\noindent
By repeating standard techniques, it is not difficult to get the expression
$$
<0\mid T\left(S(x)S(0)\right)\mid 0>=-{Z^{-1}\over
2\pi^2(x^2-i\epsilon)}{\cal K}(x)
\autoeqno{eq234}
$$
\noindent
where
$$
\eqalign{
&{\cal K}(x)=\exp\left\{-4i\pi\left[{a\over
(a-r^2)(a-r^2+1)}\left(D(x,m)-D_{1-\omega}(0,m)\right)+\right.\right.\cr
&+\left.\left.{1-r^2\over
a-r^2+1}\left(D_{1-\omega}(0,0)-D_{1-\omega}(x,0)\right)\right]\right\}\cr}
\autoeqno{eq235}
$$
\noindent
and
$$
Z=\exp\left\{4i\pi{r^2\over a-r^2}D_{1-\omega}(0,m)\right\}.
\autoeqno{eq236}
$$
Dimensional regularization is understood.

Let us now discuss the quantity $Z^{-1}{\cal K}$, which represents the
deviation from the free theory result
$$
\eqalign{
Z^{-1}{\cal K}=&\exp\left[{2a\over
(a-r^2)(a-r^2+1)}K_0\left(m\sqrt{-x^2+i\epsilon}\right)+\right. \cr
&+\left.{1-r^2\over a-r^2+1}ln\left(\tilde m^2(-x^2+i\epsilon)\right)\right]
\cr}
\autoeqno{eq239}
$$
For small values of $x^2$, we get
$$
Z^{-1}{\cal K}\sim_{x^2\rightarrow 0}\tilde C_0(-x^2+i\epsilon)^{-{r^2\over
a-r^2}},
\autoeqno{eq240}
$$
\noindent
whereas, for large negative $x^2$,
$$
Z^{-1}{\cal K}\sim_{x^2\rightarrow\infty}\tilde
C_\infty\left(-x^2+i\epsilon\right)^{{1-r^2\over a-r^2+1}},
\autoeqno{eq241}
$$
\noindent
$\tilde C_0$, $\tilde C_\infty$ being suitable constant quantities. Again
the ultraviolet behaviour can be recovered from the anomalous dimension
related to $Z$.

In the first window $(a>r^2)$, we have a singular behaviour at short
distances (negative exponent in \eqrefp{eq240}) and, since $-1+{1-r^2\over
a-r^2+1}<0$, a decreasing behaviour of the correlation function at
infinity.
We interprete this phenomenon as the existence of
a long range interaction mediated by the massless bosonic excitation.
If $r^2=1$, we see from \eqrefp{eq234}, \eqrefp{eq241} that the correlation
function of the condensate decreases at infinity as in the free theory. We
know indeed that, in this case, one of the fermions with a definite
chirality is free and the other one has only a short range interaction,
as the long range massless excitation decouples in this case.

In the second window, both exponents $-1-{r^2\over a-r^2}$ and
$-1+{1-r^2\over a-r^2+1}$ are positive. The correlation decreases at short
distances and increases when $x^2\rightarrow -\infty$. This is again a sign
of confinement.
In the correlation function for the condensates we can take first the limit
$r\rightarrow 0$ and then $a\rightarrow 0$, thereby recovering the result
we expect in the gauge invariant Schwinger model. We obtain a correlation
function which goes to a constant at infinity, as expected, since fermions
are confined in that model.
We defer the discussion concerning currents and the related charges to the
canonical treatment in the sequel.

We end this section by remarking the non trivial behaviour of this model
under a scale transformation. We notice that conformal invariance is
recovered both in the ultraviolet and in the infrared limit, with different
scale coefficients.
%This suggests the presence of an interesting renormalization group flow,
%probably related to the C--theorem \upcite{Zam86}. These aspects are presently
%under investigation.

\autosez{ope}\underbar{Operatorial approach: the bosonic sector}

\par
In this section we canonically implement the quantum dynamics of the model
described by the effective Lagrangian \eqrefp{eq27} using a Dirac--bracket
formalism \upcite{Dir50}.
Actually, this procedure only concerns the bosonic sector of the theory
\eqrefp{eq21}; nevertheless the scalar degrees of freedom will appear as the
``building blocks" in the explicit construction of a fermionic operator
solving the equations of motion derived from \eqrefp{eq21}. The possibility
of constructing fermionic operators in terms of bosonic ones (bosonization)
is a well known property of the two dimensional world \upcite{Col75}
and it turns out
to be essential in our solution and interpretation of the model.

{}From the Lagrangian \eqrefp{eq27} we obtain the momenta canonically
conjugate to the coordinates $A^0$, $A^1$ and $\phi$ (we call
$e^2/\pi=\hat e^2)$
$$
\Omega_1\equiv\Pi_0=0,
\autoeqno{eq31}
$$
$$
\Pi_1=F_{01},
\autoeqno{eq32}
$$

$$
\Pi_\phi=\partial_0\phi-\hat e rA_0-\hat e A_1,
\autoeqno{eq33}
$$
\noindent
where $\Omega_1$ is the primary constraint. The usual total Hamiltonian is:
$$
H=H_0+\int dx^1\xi_1(x^1)\, \Omega_1(x^1)
\autoeqno{eq34}
$$
\noindent
with the introduction of the Lagrange multiplier $\xi_1$ and the expression
$$
\eqalign{
H_0=\int dx^1&\left[{1\over 2}\Pi_1^2+(\partial_1 A_0)\Pi_1+{1\over
2}\Pi^2_\phi+{1\over 2}(\partial_1\phi)^2+\right.\cr
&\left.+{\hat e^2\over 2}A_0^2(r^2-a)+{1\over 2}\hat e^2(a+1)A_1^2-\hat e
r(\partial_1\phi)A_1-\right.\cr
&\left.-\hat e(\partial_1\phi)A_0+\hat e rA_0\Pi_\phi+\hat eA_1\Pi_\phi+\hat
e^2 rA_0
A_1\right],\cr}
\autoeqno{eq35}
$$
\noindent
derived from \eqrefp{eq27} by a Legendre transformation.
\noindent
Requiring that the primary constraint persists in time, we find the secondary
constraint:
$$
\Omega_2(x^1)\equiv\partial_1\Pi_1-\hat e^2(r^2-a)A_0+\hat
e\partial_1\phi-\hat e r\Pi_\phi-\hat e^2 rA_1=0.
\autoeqno{eq36}
$$
No new constraint arises for $a\not=r^2$: the Poisson bracket
$$
\left\{\Omega_1(x^1), \Omega_2(y^1)\right\}=\hat
e^2(r^2-a)\delta(x^1-y^1)
\autoeqno{eq37}
$$
\noindent
does not vanish and hence the condition $\Omega_2(x^1)=0$ only determines the
Lagrange multiplier $\xi_1(x^1)$: we are in presence of a system with
second class constraints. The discussion of the limiting case $a=r^2$ is
deferred to Appendix B.

Following the standard procedure, we introduce the Dirac bracket, derived
from \eqrefp{eq37}
$$
\eqalign{
\left\{Q(x^1), P(y^1)\right\}_D=&\left\{Q(x^1), P(y^1)\right\}-{1\over \hat
e^2(r^2-a)}\int dz^1\left[-\left\{Q(x^1),
\Omega_1(z^1)\right\}\cdot\right.\cr
&\cdot\left.\left\{\Omega_2(z^1), P(y^1)\right\}+\left\{Q(x^1),
\Omega_2(z^1)\right\}\left\{\Omega_1(z^1), P(y^1)\right\}\right],\cr}
\autoeqno{eq38}
$$
\noindent
leading to the canonical structure (we report only the non--zero brackets)
$$
\eqalign{
&\left\{A_1(x^1), \Pi_1(y^1)\right\}_D=\delta(x^1-y^1),\qquad \left\{A_0(x^1),
A_1(y^1)\right\}_D=-{1\over \hat e^2(r^2-a)}\partial_{x^1}\delta(x^1-y^1),\cr
&\left\{\phi(x^1),\Pi_\phi(y^1)\right\}_D=\delta(x^1-y^1),\qquad
\left\{A_0(x^1),
\Pi_\phi(y^1)\right\}_D={1\over \hat
e(r^2-a)}\partial_{x^1}\delta(x^1-y^1),\cr
&\left\{A_0(x^1),\Pi_1(y^1)\right\}_D=-{r\over r^2-a}\delta(x^1-y^1),\cr
&\left\{A_0(x^1), \phi(y^1)\right\}_D={r\over \hat
e(r^2-a)}\delta(x^1-y^1).\cr}
\autoeqno{eq39}
$$

In the Dirac--Bargmann formalism, the equations of motion can be written
as
$$
\dot g=\left\{g, H_{red}\right\}_D \mid_{\Omega_i=0},
\autoeqno{eq310}
$$
\noindent
$g$ being any function of canonical variables. $H_{red}$ is obtained from
$H_0$, by expressing $A_0$ as the solution of the constraint $\Omega_2=0$:
$$
\eqalign{
H_{red}&=\int dx^1\left[{1\over 2}\Pi^2_1+{1\over 2}{a\over
a-r^2}\Pi_\phi^2+{1\over 2}(\partial_1\phi)^2{a+1-r^2\over a-r^2}+\right.\cr
&\left.+{1\over 2}\hat e^2{a(a+1-r^2)\over a-r^2}A^2_1-\hat e r{a+1-r^2\over
a-r^2}A_1\partial_1\phi+\right.\cr
&\left.+\hat e{a\over a-r^2}A_1\Pi_\phi+{1\over 2}{1\over \hat e^2}{1\over
a-r^2}(\partial_1\Pi_1)^2+{1\over \hat e(a-r^2)}\cdot\right.\cr
&\left.\cdot\partial_1\phi\partial_1\Pi_1+{1\over \hat e}{r\over
a-r^2}\partial_1\Pi_1\Pi_\phi-{r\over a-r^2}A_1\partial_1\Pi_1-{r\over
a^2-r^2}\partial_1\phi\Pi_\phi\right].\cr}
\autoeqno{eq311}
$$

The quantization is now performed by taking the constraints as operatorial
equations, identifying Dirac brackets with equal time (E.T.) commutators
and using a symmetrical
ordering in the product of operators.

We remark that the breaking of gauge invariance appears in the canonical
treatment of the effective theory \eqrefp{eq27} as a change of the
constraint structure: they belong to a second class system reflecting the
absence of a local symmetry.

Using \eqrefp{eq310}, the Heisenberg equation are easily obtained and they
are completely equivalent to the Lagrange equations derived from
\eqrefp{eq27}, which was not to be ``a priori" expected
$$
\eqlabel{eq318}
\eqalignno{
\partial_\mu F^{\mu\nu}&=-\hat e^2 aA^\nu+\hat e r\partial^\nu\phi-\hat
e\tilde\partial^\nu\phi, & (\eqref{eq318}a) \cr}
$$
$$
\eqalignno{
\quadratello\phi&=r\hat e\partial_\mu A^\mu-\hat e\tilde\partial_\mu
A^\mu. &
(\eqref{eq318}b) \cr}
$$

The most general solution of these equations is
$$
\eqlabel{eq319}
\eqalignno{
A^\mu&=-{r\over a(1+a-r^2)}\partial^\mu \sigma-{(a-r^2)\over
a(1+a-r^2)}\tilde\partial^\mu \sigma+{1\over \hat e
a}(r\partial^\mu-\tilde\partial^\mu)h, & (\eqref{eq319}a) \cr}
$$
$$
\eqalignno{
\phi&=-{1\over (1+a-r^2)}\sigma+h, & (\eqref{eq319}b) \cr}
$$
\noindent
with
$$
(\quadratello+m^2)\sigma=0,
\autoeqno{eq320}
$$
$$
\quadratello\, h=0,
\autoeqno{eq321}
$$
\noindent
and $m^2$ given by \eqrefp{eq211}; $\sigma$ and $h$ describe the bosonic
degrees
of freedom of the theory. In order to show the equivalence with the
path--integral results, we are left with computing their equal--time
commutation relations, which in turn will exhibit their effective
independence and will provide us with the unitarity conditions.

{}From the identification $\sigma=\hat e\Pi_1$, we get
$$
\eqlabel{eq322}
\eqalignno{
\left[\sigma(x), \sigma(y)\right]_{E.T}&=0, & (\eqref{eq322}a) \cr}
$$
$$
\eqalignno{
\left[\sigma(x), \dot \sigma(y)\right]_{E.T.}&=i\hat e^2
m^2\delta(x^1-y^1), & (\eqref{eq322}b)
\cr}
$$
\noindent
where we have used the Heisenberg equation for $\Pi_1$
$$
\dot\Pi_1=-\hat e^2 aA_1+\hat e r\partial_1\phi-\hat e\partial_0\phi.
\autoeqno{eq323}
$$

Eq. (\eqref{eq319}b) gives the remaining commutation relations
$$
\eqlabel{eq324}
\eqlabel{eq325}
\halign{$#$\hfill & \hskip 1truecm \hfill $#$ & $#$\hfill &
\hskip 0.6truecm \hfill $#$ \cr
\left[h(x), \sigma(y)\right]_{E.T.}=0, & (\eqref{eq324}a) \quad&
\left[h(x), h(y)\right]_{E.T.}=0, & (\eqref{eq325}a) \cr
\noalign{\vskip 0.5truecm}
\left[h(x), \dot \sigma(y)\right]_{E.T.}=0, & (\eqref{eq324}b) \quad&
\left[h(x), \dot h(y)\right]_{E.T.}=i{a\over 1+a-r^2}
\delta(x^1-y^1). & (\eqref{eq325}b)\cr}
$$

In particular eqs. \eqrefp{eq324} show the indipendence of
massive and massless degrees of freedom. The request of the absence of
tachyons from the spectrum forces the parameters $a$ and $r$ to range in
the two windows \eqrefp{eq216}.

In the first one $(a>r^2)$ the commutation
relations (\eqref{eq322}b) and (\eqref{eq325}) are physical, so that
$\sigma$ and
$h$ generate a Fock space with a positive defined metric.
\noindent
We remark that, from a rigorous point of view, the positivity of the
massless sector is achieved only after a Krein extension of the original
Fock topology derived from (\eqref{eq319}b) \upcite{Mor90};
the realization of such
non--trivial extension is also essential in order to prove the existence of
the operators that, in the next section, we will construct to describe the
fermionic degrees of freedom of the theory.

In the other window $(0<a<r^2-1~$ or $~ r^2-1<a<0)$ $h$ is a ``ghost", having
the negative sign in its commutation relations.
We can define a physical Hilbert space imposing the subsidiary condition
$$
h^+(x)\mid\Phi_{phys}>=0,
\autoeqno{eq326}
$$
\noindent
which however possesses a non local character with respect to $A_\mu$.
\noindent
This condition is stable under time evolution, due to the free character of
$h$. Obviously \eqrefp{eq326} selects the physical operators of the theory:
in other words it imposes a restriction on the operators representing the
fermionic sector, as we will see in the next section.

Now we try to discuss some limiting situations on the parameters $a$ and
$r$, but the case $a=r^2$ that involves a doubling of the constraints and is
deferred to Appendix B.

The commutation relations \eqrefp{eq325} are singular in the limit
$a=r^2-1$;
nevertheless, if we come back to equations of motion \eqrefp{eq319} and we
put $a=r^2-1$, we can solve for $A_\mu$ and $\phi$ without the occurence of
any singularity.
The solution is
$$
A_\mu=\hat e {1\over \quadratello}
(\tilde\partial_\mu-r\partial_\mu)\sigma-{1\over \hat
e(r^2-1)}\tilde\partial_\mu \sigma+{r\over \hat e}\tilde\partial_\mu h+{1\over
\hat e(r^2-1)}\partial_\mu h,
\autoeqno{eq327}
$$
$$
\phi=\hat e^2(1-r^2){1\over \quadratello}\sigma+h,
\autoeqno{eq328}
$$
\noindent
where
$$
\quadratello\, \sigma=\quadratello\, h=0
\autoeqno{eq329}
$$
\noindent
and ${1\over \quadratello}$ is the inverse of the d'Alembert operator.
Clearly the relation among $A_\mu, \phi$ and $\sigma$ is not local (due to the
presence of an integral operator) and the theory seems to lose its local
character.
The other limiting case is $a=0~(r\not=0)$: this limit corresponds to a
``would be" gauge invariant regularization of the theory, and it can be
performed starting from the second window.
The mass vanishes and we recognize a situation similar to the one in the
case $a=r^2-1$
$$
\eqalign{
&A_\mu={1\over \hat e}{1\over \quadratello}\partial^\mu \sigma-\hat e
{1\over
\quadratello}\tilde\partial^\mu \sigma+h^\mu_1,\cr
&\quadratello\, \sigma=0,\cr
&\quadratello\, h_1^\mu=0, \quad
(r\partial_\mu-\tilde\partial_\mu)h_1^\mu=0,\cr}
\autoeqno{eq330}
$$
\noindent
while
$$
\quadratello\,\phi =0.
\autoeqno{eq341}
$$

Again the properties of the theory are not transparent, due to the non
local relation with the basic degrees of freedom.
The situation is reminiscent of the chiral Schwinger model for $a=0$
studied in \upcite{Gir86}. We observe that it is possible to put $r=0$: it
would
correspond to having a Schwinger model regularized in a gauge dependent
way.

For $a>0$ we have two positive metric field: $\sigma$ with mass $\hat e^2(1+a)$
and $h$ massless; for $-1<a<0$, $h$ becomes a ``ghost".
These theories are not equivalent to the original Schwinger model: the
introduction of the gauge--breaking counterterm
$$
{\hat e a\over 2}A_\mu A^\mu
\autoeqno{eq342}
$$
\noindent
cannot be interpreted as a gauge fixing and the model rather resembles to
the Stuckelberg electrodynamics in 2 dimensions.

In conclusion we have recovered in an operatorial formalism,
the results of
the path--integral approach, concerning the bosonic sector.
In particular the propagator
$$
<0\mid T(A_\mu(x)A_\nu(y)\mid 0>
$$
\noindent
can be computed and coincides with \eqrefp{eq210}, apart from irrilevant
non--covariant contact terms. Moreover the structure of the Hilbert space
of states has been clarified in the various cases.

A last remark concerns the singularity of the solutions when $\hat
e\rightarrow 0$: our results are truly non perturbative as we do not
introduce any gauge fixing which would be necessary to build a free propagator
to start with.

\autosez{fer}\underbar{Operatorial approach: the fermionic sector}

\par
In order to establish a definitive link with the path--integral formalism,
we have to construct the fermionic operator that solves the equations
derived from \eqrefp{eq21}. Actually we will go further, finding a
conserved charge that allows us to identify a fermionic sector on the
Hilbert space of the model: in this way the difference between the two
windows \eqrefp{eq216} will be fully enlightened, confirming the analysis
of the path--integral formulation.

Let's come back to the original Lagrangian \eqrefp{eq21} and obtain the
Maxwell and Dirac equations
$$
\partial_\mu F^{\mu \nu}=-e J^\nu,
\autoeqno{eq41}
$$
$$
i \parz\psi+ e \A(1+r\gamma^5)\psi=0 ,
\autoeqno{eq42}
$$
\noindent
with the ``classical" current $J^\mu$ defined as
$$
J^\mu=\bar\psi\gamma^\mu(1+r\gamma^5)\psi.
\autoeqno{eq43}
$$

In solving these equations we need a regularization procedure to give a
meaning to the composite operators $\A\psi(x)$ and $J^\mu(x)$: we seek
consistency with the results of the bosonic sector. In so doing we are able
to express $\psi$ as a well defined functional of the bosonic
degrees of freedom $\sigma$ and $h$.

Taking the expression \eqrefp{eq319}a of $A_\mu$ into account, it is easy
to verify that a classical solution of \eqrefp{eq42} is
$$
\psi(x)=\exp{i\sqrt{\pi}\over a}\left[-\left(r+{a\over
1+a-r^2}\gamma^5\right)\sigma(x)+ \left(r^2-1\right)\gamma^5
h(x)\right]\psi_0(x),
\autoeqno{eq44}
$$
\noindent
where $\psi_0(x)$ obeys to the free Dirac equation. To obtain an operator
solution, we define $\A\psi(x)$ by normal ordering $:\A\psi:(x)$  and
use the bosonized form of $\psi_0(x)$ \upcite{Col75}; we get
$$
\eqalign{
\psi_\alpha(x)&=C\sqrt{{\mu\over 2\pi}}:\exp{i\sqrt{\pi}\over
a}\left[-\left(r+{a\over
1+a-r^2}\gamma^5_{\alpha\alpha}\right)\sigma\right. \cr
&\left.+\left(r^2-1\right)\gamma^5_{\alpha\alpha}h+a\varphi-a
\gamma^5_{\alpha\alpha}\tilde\varphi\right]: ,\cr}
\autoeqno{eq45}
$$
\noindent
where $\varphi(x)$ is a massless scalar field and $\tilde\varphi$ its dual
$$
\tilde\partial_\mu\varphi=\partial_\mu\tilde\varphi,
\autoeqno{eq46}
$$
\noindent
$\mu$ is an infrared regulator associated to $\varphi$, carrying the correct
balance of canonical dimension and $C$ a normalization constant to be
determined later on. We notice that the normalization of $\varphi$ is not
fixed ``a priori" and will be suggested by the solution; moreover we have
written the second member of \eqrefp{eq45} as a single normal ordering.
This choice will turn out to be the correct one because we shall find that
$\varphi$ is proportional to $h$: there is no way of separating in the
general case $a \neq r^2$ the free contribution to the fermionic
solution from the interaction one.

The relation between $\varphi$ and $h$ , as well as the determination of the
coefficient $C$, rely on the solution of \eqrefp{eq41}.
We know from \eqrefp{eq318}a that
$$
\partial_\mu F^{\mu\nu}=\hat e\tilde\partial^\nu\sigma ;
\autoeqno{eq47}
$$
\noindent
we define a quantum current $\hat J_\mu(x)$ by a normal product $N$ to
be specified below
$$
\eqalign{
\hat J^\mu(x)=&N\left[\bar\psi\gamma^\mu(1+r\gamma^5)\psi\right]=
\left(1+r\right)
N_R\left[\bar\psi\gamma^\mu
P_R\psi\right]+\left(1-r\right)N_L\left[\bar\psi\gamma^\mu
P_L\psi\right], \cr
P_{R,L}=&{1\pm\gamma^5\over 2}, \cr}
\autoeqno{eq48}
$$
\noindent
in a way consistent with \eqrefp{eq47}: an additional request is the
infrared finiteness of such a current, that will fix the constant $C$.

Generalizing the standard point--splitting procedure, with
$\epsilon^2 < 0$,
we get
$$
N_{R,L}\left[\bar\psi\gamma^\mu
P_{R,L}\psi\right](x)=J^\mu_{R,L}(x)+a_{R,L}A^\mu(x),
\autoeqno{eq49}
$$
$$
\eqalign{
J^\mu_{R,L}(x)=&\lim_{\epsilon\rightarrow
0}U^{-1}_{R,L}(\epsilon)\left\{\bar\psi(x+\epsilon)\gamma^\mu
P_{R,L}. \right.\cr
&\left.\exp\left(i\sqrt{\pi}\hat e\int_x^{x+\epsilon}dz_\nu\left[K_1 A^\nu+
K_2 \tilde A^\nu\right]\right)\psi(x)-V.E.V.\right\},\cr}
\autoeqno{eq410}
$$
\noindent
where $V. E. V.$ stands for ``vacuum expectation value",
$U_{R,L}(\epsilon)$ are
some ultraviolet renormalization constants and $K_1, K_2, a_{R,L}$ are
numerical factors suitably choosen in order to satisfy the Maxwell
equation. In particular the presence of $a_{R,L}$ is linked to the loss of
gauge invariance: they represent the arbitrariness of the regularization up
to finite terms (not fixed by gauge invariance) and make the current
$\hat J^\mu(x)$ conserved, as requested from the Gauss' law.

We begin by considering the zero component of $N_R\left[\bar\psi\gamma^\mu
P_R\psi\right]$:
$$
\eqalign{
N_R\left[\psi ^\dagger P_R\psi\right](x)\simeq&
U^{-1}_R(\epsilon)\left\{\psi^\dagger (x+\epsilon)P_R\psi(x)+i\sqrt{\pi}\hat
e\psi^\dagger (x+\epsilon)P_R\psi(x)\epsilon^\mu\right.\cr
&\left.\left(K_1 A_\mu+K_2\tilde A_\mu\right)
+{\cal O}(\epsilon^2)-V. E. V.\right\}+a_R A^0(x).\cr}
\autoeqno{eq411}
$$

Let's suppose for the moment that $\varphi(x)$ is indipendent from $\sigma$
and $h$, and has the normalization $\rho$
$$
\left[\varphi(x),\dot \varphi(y)\right]_{E.T.}=i\rho\delta(x^1-y^1).
\autoeqno{eq412}
$$
Standard Wick's techniques lead to
$$
\eqalign{
&\psi^\dagger (x+\epsilon)P_R\psi(x)={\mu\over 2\pi}C^2:\exp-{i\sqrt{\pi}\over
a}\epsilon^\mu\partial_\mu\left[-\left(r+{a\over
a+1-r^2}\right)\sigma+\right.\cr
&\left.+\left(r^2-1\right)h+a\varphi-a\tilde\varphi\right]:
\exp{\pi\over a^2}\left\{\left(r+{a\over a+1-r^2}\right)^2 a{(a+1-r^2)\over
a-r^2}D^+(\epsilon,m)+\right.\cr
&\left.+ (r^2-1)^2
{a\over a+1-r^2}D^+(\epsilon,\mu)
+2a^2\rho D^+(\epsilon,\mu)-2a^2\rho \tilde D^+(\epsilon)\right\}\cr}
\autoeqno{eq413}
$$
\noindent
with
$$
\eqalign{
&D^+(x,m)={1\over 2\pi} K_0(m \sqrt{-x^2+i x^0 \delta}), \cr
&D^+(\epsilon,\mu)=-{1\over 4\pi}ln(-\mu^2 \epsilon^2+i x^0 \delta) ,\cr
&\tilde D^+(\epsilon)={1\over 4\pi}\left[ln(\epsilon^{-}-i\delta)-
ln(\epsilon^{+}-i\delta)\right],\cr
&\epsilon^{\pm}=\epsilon^0\pm \epsilon^1,\qquad \delta>0. \cr}
\autoeqno{eq414}
$$

We notice that the left term of the equality \eqrefp{eq413} is the sum of
the $0$ and $1$ components of the two--vector $\bar\psi\gamma^\mu\psi$, while
in the second member we have only scalar quantities, but
$\tilde D^+(\epsilon)$;
we fix $\rho$ to recover the correct tensorial structure. The relevant
term is:
$$
\exp{\pi\over a^2}\left[2a^2\rho D^+(\epsilon,\mu)-2a^2\rho\tilde
D^+(\epsilon)\right]=\exp\left[-{i\pi\rho\over 2}-\rho
ln\mu\right]\left({\epsilon^+\over \epsilon^2}\right)^\rho,
\autoeqno{eq415}
$$
\noindent
forcing $\rho=1$. Then we choose
$$
C=\left({\mu\over \tilde m}\right)^{{1\over 4}{(r^2-1)^2\over a(a+1-r^2)}}
\autoeqno{eq416}
$$
\noindent
in order to obtain a result independent of the infrared regulator
$$
\eqalign{
\left[\psi^\dagger (x+\epsilon)P_R\psi(x)-V. E.
V.\right]=&U_R(\epsilon)\left\{{1\over 2\sqrt{\pi}a}{\epsilon^\mu\over
\epsilon^{-}}\partial_\mu\left[\left(r+{a\over
a+1-r^2}\right)\sigma+\right.\right.\cr
&\left.\left.+(1-r^2)h\right]-
-{1\over 2\sqrt{\pi}}{\epsilon^\mu\over
\epsilon^{-}}(\partial_\mu-\tilde\partial_\mu)\varphi+0(\epsilon)\right\},\cr}
\autoeqno{eq417}
$$
$$
\eqalign{
U_R(\epsilon)=&\exp{\pi\over a^2}\left\{\left(r+{a\over
a+1-r^2}\right)^2{a(a+1-r^2)\over a-r^2}D^+(\epsilon,m)\right.\cr
&\left.-{1\over
4\pi}{a(r^2-1)^2\over a+1-r^2}ln(-\tilde
m^2\epsilon^2+i\epsilon^0\delta)\right\}.\cr}
\autoeqno{eq418}
$$

One should compare this expression with the fermion wave function
renormalization constant in \eqrefp{eq226}. Apart from the different
regularization, they manifestly exhibit the same behaviour
$U_R(\epsilon) \sim Z_R^{-1}$, which is rooted in the fact that
the current $J_R^{\mu}$ does not undergo renormalization.

In the same way we get
$$
\eqalign{
&i\sqrt{\pi}\hat e\psi^\dagger (x+\epsilon)P_R\psi(x)\epsilon^\mu(K_1
A_\mu+K_2\tilde A_\mu)-V. E. V. = \cr
& =U_R(\epsilon)\left\{{\hat e\over
2\sqrt{\pi}}{\epsilon^\mu\over \epsilon^{-}}(K_1 A_\mu+K_2\tilde
A_\mu)+0(\epsilon)\right\}. \cr}
\autoeqno{eq419}
$$

Collecting all terms, we end up with
$$
\eqalign{
&N_R\left[\psi^\dagger (x)P_R\psi(x)\right]=\lim_{\epsilon\rightarrow
0}\left\{-{1\over 2\sqrt{\pi}}{\epsilon^\mu\over
\epsilon^-}(\partial_\mu-\tilde \partial_\mu)\varphi\right.+\cr
&\left.+{1\over 2\sqrt{\pi}}{\epsilon^\mu\over \epsilon^-}{1\over
a}\left[\partial_\mu\sigma\left(r+{a\over 1+a-r^2}-{r\over
1+a-r^2}K_1-{a-r^2\over 1+a-r^2}K_2\right)\right.\right.-\cr
&\left.\left.-\tilde\partial_\mu \sigma\left(K_1{a-r^2\over 1+a-r^2}+K_2{r\over
1+a-r^2}\right)+\left(1-r^2+rK_1-K_2\right)\partial_\mu h\right.\right.+\cr
&\left.\left.+\left(rK_2-K_1\right)\tilde\partial_\mu h\right]\right\}+a_R
A^0(x).\cr}
\autoeqno{eq420}
$$

It is quite natural to put $K_1=1$; $K_2=r$ to recover a
direction independent limit as $\epsilon\rightarrow 0$:
$$
\eqalign{
N_R\left[\psi^\dagger (x)P_R\psi(x)\right]=&-{1\over
2\sqrt{\pi}}(\partial_0-\partial_1)\varphi+{1\over 2\sqrt{\pi}a}\left\{{a\over
a+1-r^2}(\partial_0-\partial_1)\sigma \right.\cr
&\left.+(1-r^2)(\partial_0-\partial_1)
h\right\}+a_R A^0(x).\cr}
\autoeqno{eq421}
$$

{}From this expression, using $\gamma^0 P_R=\gamma^1 P_R$, we
immediately reconstruct
$$
\eqlabel{eq422}
\eqalignno{
J^\mu_R&={1\over
2\sqrt{\pi}}(\tilde\partial^\mu-\partial^\mu)\left[\varphi-{1\over
a+1-r^2}\sigma-{1-r^2\over a}h\right],&(\eqref{eq422}a)
\cr}
$$
$$
\eqalignno{
J^\mu_L&=-{1\over
2\sqrt{\pi}}(\tilde\partial^\mu+\partial^\mu)\left[\varphi+{1\over
a+1-r^2}\sigma+{(1-r^2)\over a}h\right].&(\eqref{eq422}b)
\cr}
$$

As we have predicted, these currents are not conserved: requiring the
consistency of the Maxwell equation and setting $a_{R,L}={a\hat e\over
2\sqrt{\pi}}\alpha_{R,L}$, we obtain
$$
(1+r)\alpha_R+(1-r)\alpha_L=2.
\autoeqno{eq423}
$$

We can choose
$\alpha_R=\alpha_L=1$;
a relation is thereby induced between $\varphi$ and $\tilde h$
$$
\varphi=-{a+1-r^2\over a}\tilde h
\autoeqno{eq424}
$$
\noindent
and the Maxwell equation can be rewritten as
$$
\partial_\mu F^{\mu\nu}=-\sqrt{\pi}\hat e\hat J^\nu=-\hat
e\sqrt{\pi}\left\{(1+r) J^\nu_R+(1-r) J^\nu_L+{a\hat e\over
2\sqrt{\pi}}A^\nu\right\}.
\autoeqno{eq425}
$$

Eq.\eqrefp{eq424} contradicts our initial hypothesis of independence of
$\varphi$
from $h$: hence we are forced to come back to \eqrefp{eq411} and to impose
$$
\varphi=b\tilde h,\qquad b\in\reali,
$$
\noindent
getting
$$
\rho=b^2{a\over a+1-r^2}.
$$

If we repeat the calculation taking into account the new commutator
$[\varphi,
h]$, the equation to solve in order to recover the correct tensorial
structure is
$$
{a\over a+1-r^2}b^2-{r^2-1\over a+1-r^2}b-1=0,
\autoeqno{eq426}
$$
\noindent
leading to the roots:
$$
b_1=-{a+1-r^2\over a},\qquad b_2=1.
\autoeqno{eq427}
$$

Now all the equations from \eqrefp{eq416} to \eqrefp{eq424} hold true
since the dependence on the normalization of $\varphi$ is encoded in the
exponent of ${\epsilon^+ \over \epsilon^2}$, that is always unity.
Then Maxwell consistency selects $b=b_1$.

In conclusion the fermionic operator satisfying the equations of motions
\eqrefp{eq41}, \eqrefp{eq42} is
$$
\eqalign{
\psi_\alpha(x)=&C\sqrt{{\mu\over 2\pi}}:\exp-{i\sqrt{\pi}\over
a}\left[\left(r+{a\over a+1-r^2}\gamma^5_{\alpha\alpha}\right)\sigma+
\right.\cr
&\left.+\left(a+1-r^2\right)\tilde h-a\gamma^5_{\alpha\alpha}h\right]:.
\cr}
\autoeqno{eq428}
$$

We can recover the known results for the chiral Schwinger model putting
$r=1$
$$
\psi_R=\psi^0_R(x),\qquad
\psi_L(x)=:\exp-2i\sqrt{\pi}\sigma:\psi^0_L(x).
\autoeqno{eq429}
$$

In this case the interacting solution factorizes into an interaction piece,
depending on the massive component, and a free fermion $\psi^0(x)$; its
asymptotic behaviour is the one of a free Dirac theory. We remark that only
for $r=\pm 1$ the free part is factorized.

One can easily check that the Green functions
$$
<0\mid T\left(\psi_\alpha(x)\psi^\dagger_\beta(y)\right)\mid 0>,\qquad
<0\mid T\left(\bar\psi\psi(x)\bar\psi\psi(y)\right)0>,
$$
computed using \eqrefp{eq428}, coincide with the path--integral
ones \eqrefp{eq224}, \eqrefp{eq234}, apart from the wave--function
renormalization constant for the field $\psi$ (our solution is here
renormalized).

Now we want to discuss the properties of the solution \eqrefp{eq428}. As
starting point we remark that gauge invariance is completely broken; hence
$\psi_\alpha(x)$ is not affected, in principle, by any gauge ambiguity.

The
first investigation concerns the electric charge of this solution:
integrating the zero component of the conserved current $\hat J_\mu(x)$
(that couples to the Gauss's law), we get the generator
$$
\hat Q=\int dx^1 \hat J_0(x^1).
\autoeqno{eq430}
$$

A simple calculation gives the commutation rule
$$
\left[\hat Q, \psi_\alpha(x)\right]=0,
\autoeqno{eq431}
$$
\noindent
showing that $\psi_\alpha(x)$ is electrically neutral; actually, in order
to be rigorous, one should smear $\hat Q$ with a test function of
compact support $f_R$ and prove that
$$
\lim_{R\to \infty}\left[\hat Q_R, \psi_\alpha(x)\right]=0
$$
\noindent
with $\hat Q_R=\int dx^1 \hat J_0(x^1)f_R(x^1) $.

The electric charge of the original fermion is totally screened: this is
true for any value of $r$ and $a$.

At this point we recall that, in the first window $(a>r^2)$, $\psi$ is a well
defined operator on the Hilbert space of $\sigma$ and $h$
with the prescription of taking the limit $\mu \rightarrow 0$ on its
correlation
functions; moreover $\psi$ generates a positive norm Hilbert space,
whose properties will be specified in the next section.

On the other
hand, in the second window, we have to impose on the physical operators the
condition \eqrefp{eq320} equivalent to
$$
\left[h^+(x), \Phi_{phys}\right]=0.
\autoeqno{eq432}
$$
A short calculation:
$$
\eqalign{
&\left[h^+(x), \psi_\alpha(y)\right]=-{i\sqrt{\pi}\over a}\left[h^+(x),
(a+1-r^2)\tilde
h^-(y)-a\gamma^5_{\alpha\alpha}h^-(y)\right]\psi_\alpha(y)\cr
=&-{i\sqrt{\pi}\over a}\psi_\alpha(y)\left[a\left(\tilde D^+(x-y)-{i\over
4}\right)-{a^2\over
a+1-r^2}\gamma^5_{\alpha\alpha}D^+(x-y,\mu)\right]\not=0,\cr}
\autoeqno{eq433}
$$
shows that $\psi_\alpha(x)$ fails to be physical.
%In other words we cannot
%construct a fermionic operator, solving the model and living in a positive
%norm Hilbert space. We conclude that in the second window the spectrum of
%the theory consists only of massive bosons.
This analysis agrees  with the
path--integral one and is confirmed by the inspection of the (anti)
commutation relations.
We study
$$
\{\psi_\alpha(x), \psi^\dagger_\beta(0)\}_{E.T.}.
\autoeqno{eq434}
$$
\noindent
For $\alpha\not=\beta$ is straightforward to show (using the standard
properties of Green functions in $1+1$ dimensions), that the result is
zero. For $\alpha=\beta$ the computation gives
$$
\eqalign{
&\left\{\psi_\alpha(x), \psi^\dagger_\alpha(0)\right\}_{E.T.}=:\exp
\left\{-{i\sqrt{\pi}\over a}\left[\left(r+{a\over
a+1-r^2}\gamma^5_{\alpha\alpha}\right)\sigma(x)+(a+1-r^2)\tilde
h(x)\right.\right.-\cr
&\left.\left.- a\gamma^5_{\alpha\alpha}h
-\left(r+{a\over
a+1-r^2}\gamma^5_{\alpha\alpha}\right)\sigma(0)-\left(a+1-r^2\right)\tilde
h(0)+a\gamma^5_{\alpha\alpha}h(0)\right]\right\}:\cr
&\cdot A_{\alpha\alpha}(x)B_{\alpha\alpha}(x), \cr}
\autoeqno{eq435}
$$
\noindent
where
$$
\eqlabel{eq436}
\eqalignno{
A_{\alpha\alpha}(x)&=C^2\exp{\pi\over a^2}\left[\left(r+{a\over
a+1-r^2}\gamma^{\alpha\alpha}_5\right)^2{(a+1-r^2)a\over
a-r^2}D^+(x,m)\right.+&\cr
&\left.+{a(1-r^2)^2\over a+1-r^2}D^+(x,\mu)\right], & (\eqref{eq436}a)\cr
B_{\alpha\alpha}(x)&={\mu\over
2\pi}\exp2\pi\left[\left(D^+(x,\mu)-\gamma^5_{\alpha\alpha}\tilde
D^+(x)\right)+\left(D^+(-x,\mu)-\gamma^5_{\alpha\alpha}\tilde
D^+(-x)\right)\right].& \cr
&  & (\eqref{eq436}b)\cr}
$$

For $x^0=0$ we get $B_{\alpha\alpha}(x)=\delta(x^1)$, so that
$$
\eqalign{
&\left\{\psi_\alpha(x),
\psi^\dagger_\alpha(0)\right\}_{E.T.}=A_{\alpha\alpha}(0)\delta(x^1), \cr
&A_{\alpha\alpha}(0)=\exp\pi{(r+\gamma^5_{\alpha\alpha})^2\over a-r^2}
D_{1-\omega}(0,m).\cr}
\autoeqno{eq437}
$$

Recalling \eqrefp{eq226},we find
$$
\eqlabel{eq438}
\eqalignno{
A_{11}=&Z^{-1}_R,&(\eqref{eq438}a) \cr
A_{22}=&Z^{-1}_L.&(\eqref{eq438}b) \cr}
$$

Eq.\eqrefp{eq437} are anticommutation relations
for interacting fermions
(see e. g. \upcite{Itz85}).

In the same way
$$
\left\{\psi_\alpha(x), \psi_\beta(0)\right\}_{E.T.}=0 \qquad\qquad
{}~\forall_{\alpha,\beta}.
$$

In the next section we shall restrict ourselves to the case $a>r^2$, where
we shall succeed in giving a deeper characterization of the solution
in this case.

\autosez{rel}\underbar{The relation with the massless Thirring model}

\par
We have seen that for $r^2=1$ the solution \eqrefp{eq428} factorizes into
an interaction piece depending on the massive boson $\sigma$ and a free
spinor: the asymptotic behaviour of the Green functions is the one of free
chiral fermions. The solution is electrically neutral and carries the
fermion number associated to the free conserved current \upcite{Gir86}
$ J_0^\mu(x)=\bar\psi_0\gamma^\mu \psi_0(x)$

$$
\hat Q^{(0)}=\int dx^1 J_0^{(0)}(x^1), \qquad\qquad \left[\hat
Q^{(0)},\psi_\alpha(x)\right]=\psi_\alpha(x).
\autoeqno{eq51}
$$

The conclusion is that a free massless fermion exists as asymptotic state.

In the general situation $r^2\not=1$ , as we have seen, we cannot draw a
similar conclusion, due to the long range character of the interaction.
Nevertheless a solution of the Dirac equation exists, carrying the correct
anticommutation relation: we try to find what kind of states are linked
to this operator.
\noindent
Due to the indipendence of $\sigma$ and $h$ we can factorize $\psi_\alpha$
as
$$
\eqalign{
\psi_\alpha(x)=C\sqrt{{\mu\over 2\pi}}:\exp &-{i\sqrt{\pi}\over
a}\left[\left(r+{a\over
a+1-r^2}\gamma^5_{\alpha\alpha}\right)\sigma\right]:\cr
:\exp &-{i\sqrt{\pi}\over a}\left[\left(a+1-r^2\right)\tilde
h-a\gamma^5_{\alpha\alpha}h\right]:.\cr}
\autoeqno{eq52}
$$

First we look at the ``spin" of this operator: we study the transformation
property under Lorentz boost of the correlation function
$$
\eqalign{
<0\mid\psi_\alpha(x)\psi^\dagger_\alpha(0)\mid 0>=C^2 &\exp\left\{\pi
m^2\left[{r\over a}+{1\over
a+1-r^2}\gamma^5_{\alpha\alpha}\right]D^+(x,m)\right\}\cdot\cr
\cdot &\exp\left\{\pi\left({a+1-r^2\over a}+{a\over
a+1-r^2}\right)D^+(x,\mu)\right\}\cdot \cr
&\cdot \exp\left(-2\pi\gamma^{\alpha\alpha}_5\tilde
D^+(x)\right).\cr}
\autoeqno{eq53}
$$

Calling $\chi$ the parameter of the Lorentz boost $\sinh \chi={v\over
\sqrt{1-v^2}}$, the transformation of the massless commutators
$D^+(x,\mu)$
and $\tilde D^+(x)$ are
easily found to be
$$
\eqlabel{eq54}
\eqalignno{
D^+(x,\mu)&\rightarrow D^+(x,\mu), &(\eqref{eq54}a)
\cr}
$$
$$
\eqalignno{
\tilde D^+(x)&\rightarrow \tilde D^+(x)-{\chi\over 2\pi}. &(\eqref{eq54}b)
\cr}
$$

The boost on \eqrefp{eq53} acts as
$$
<\psi_\alpha(x)\psi^\dagger_\alpha(0)>\rightarrow
<\psi_\alpha(x)\psi^\dagger_\alpha(0)>\exp(\gamma^5_{\alpha\alpha}\chi),
\autoeqno{eq55}
$$
\noindent
that suggests the rule
$$
\psi(x)\rightarrow \exp\left({1\over 2}\gamma^5\chi\right)\, \psi(x).
\autoeqno{eq56}
$$

The ``spin" is $s={1\over 2}$ (independent of $r$ and $a$); we remark we
are not talking about a true spin, as no rotation group is present in two
dimensions. Hence the ``spin" is rather a label for the representation of the
Lorentz group.

Then we turn our attention to scaling properties: the question is subtler
because the existence of the field $\sigma$. The explicit presence of a mass
violates scale invariance: in the limit $x^2\rightarrow +\infty$, when
the massive components decouple from the correlation function, we can
recover an exact scaling. It is not difficult to read the asymptotic scale
dimension of $\psi_\alpha(x)$ from \eqrefp{eq53}, in this limit. Under a
dilatation $x_\mu\rightarrow \lambda x_\mu$
$$
\eqlabel{eq57}
\eqalignno{
D^+(x,\mu)&\rightarrow D^+(x,\mu)-{\lambda\over 2\pi}, &(\eqref{eq57}a)
\cr}
$$
$$
\eqalignno{
\tilde D^+(x)&\rightarrow \tilde D^+(x), &(\eqref{eq57}b)
\cr}
$$
\noindent
giving
$$
\psi_\alpha(x)\rightarrow \psi_\alpha(x)\exp\left( -\lambda{1\over
4}\left[\left(1+g\right)+{1\over 1+g}\right]\right),
\autoeqno{eq58}
$$
\noindent
where
$$
g={1-r^2\over a}.
\autoeqno{eq59}
$$

The asymptotic scale dimension (that we identify with the scale dimension
of the asymptotic state) is
$$
d={1\over 4}\left[\left(1+g\right)+{1\over 1+g}\right].
\autoeqno{eq510}
$$
Obviously this result is fully consistent with the analysis of the
anomalous
dimension of the propagator for $x^2\rightarrow -\infty$ \eqrefp{eq230};
using the notation of sect.$2$ , we get
$$
d={1\over 2}+B.
$$

We notice that for $g=0$ we recover the free spinor of the chiral Schwinger
model : in a precise sense, that we discuss below, $g$ describes a kind of
asymptotic interaction.

The propagator \eqrefp{eq230} in the large $x$ limit is the propagator of
the massless Thirring model \upcite{Kla67},
in the spin ${1\over 2}$ representation.
Actually for this model the spin labels the representation of the conformal
group \upcite{Swi75}. Our
asymptotic state is a massless Thirring fermion: we can write
$$
\psi_\alpha(x)=:\exp\left( -i\sqrt{\pi}\left({r\over a}+{1\over
a+1-r^2}\gamma_{\alpha\alpha}^5\right)\sigma\right):\hat\psi_\alpha(x),
\autoeqno{eq511}
$$
\noindent
with
$$
\eqalign{
\hat\psi_\alpha(x)=&C:\exp\left(-i\sqrt{\pi}\left[\left(1+g\right)\tilde
h-\gamma^5_{\alpha\alpha}h\right]\right):(x),\cr
&C=\sqrt{{\mu\over 2\pi}}\left({\mu\over \tilde m}\right)^{{g^2\over
a(1+g)}}.\cr}
\autoeqno{eq512}
$$

It is not difficult to show that, from the operatorial point of view,
$\hat\psi_\alpha$ is a solution of the massless Thirring model,
namely of the equation
$$
i\gamma^\mu\partial_\mu\hat\psi=\hat g:\gamma^\mu\hat J_\mu\hat\psi:,
\autoeqno{eq513}
$$
\noindent
where we have defined \upcite{Abd91}

$$
\eqalign{
&\hat J^0=\lim_{\epsilon\rightarrow 0}
Z(\epsilon)\left\{J^0(x,\epsilon)-<0\mid J^0
(x,\epsilon)\mid 0>\right\},\cr
&\hat J^1=\lim_{\epsilon\rightarrow
0}Z(\epsilon){1 \over 1+g}\left\{J^1(x,\epsilon)-<0\mid J^1
(x,\epsilon)\mid 0>\right\},\cr
&J_{\pm}(x,\epsilon)=\psi^\dagger(x,\epsilon)(1\pm\gamma^5)\psi(x),\cr
&Z(\epsilon)=(-\tilde m^2\epsilon^2)^{g^{2}\over 4(1+g)} \cr}
\autoeqno{eq514}
$$
\noindent
and
$$
\hat g=\pi g=\pi{(1-r^2)\over a}.
\autoeqno{eq515}
$$

The coupling constant of this ``effective" Thirring model depends on $r$
and $a$: for $a>r^2$ we have a dynamical generation of the Thirring theory.
One can also check directly that \eqrefp{eq512} is a Thirring fermion (spin
${1\over 2}$) looking at the Klaiber manifold
\upcite{Kla67}: eq. \eqrefp{eq510} is the
correct dimension for the spin ${1\over 2}$ solution.

We can now define the charge associated to this model
$$
\eqalign{
&\hat Q_T=\int dx^1\hat J_0(x^1),\cr
&\hat J_\mu(x)=-{1\over 2\sqrt{\pi}}\partial_\mu h(x).\cr}
\autoeqno{eq516}
$$
\noindent
$\hat Q_T$ is obviously conserved and it results
$$
\left[\hat Q_T, \psi_\alpha(x)\right]=\psi_\alpha(x).
\autoeqno{eq517}
$$
\noindent
In other words the solution $\psi_\alpha$ carries the quantum number of a
Thirring fermion.

The selection rules
are obtained setting $\mu\rightarrow 0$ in the correlation function.
This ``thermodynamic limit" is essential in order to recover the
symmetries of the original theory; for example the naive definition
$$
<0\mid \psi_\alpha\mid 0> = C\sqrt{{\mu\over 2\pi}} \neq 0
$$
suggests the spontaneous breaking of the $U(1)$ rigid symmetry generated
by $\hat Q_T$. The vacuum is not invariant under this transformation:
only when $\mu \rightarrow 0$ we recover the correct invariance. This
procedure leads to selection rules equivalent to Klaiber's ones and
ensure the positivity of the Hilbert space.

At this point we remark that all our constructions are justified, from a
rigorous mathematical point of view, by the fact that we can make a Krein
extention of the original massless boson Hilbert space in order to obtain a
representation of the fermionic algebra solving the massless Thirring model
\upcite{Pie92}. Using this technique one can define the charge operator $\hat
Q_T$ and
prove the existence of \eqrefp{eq512} in a strong operatorial sense.

The invariance of the vacuum, in this formalism, is not achieved
by means of the {\it ad hoc} infrared limit $\mu \rightarrow 0$,
but by a careful construction of the fermionic vacuum in the Krein
topology: uniqueness is obtained modulo zero norm vectors (that are
quotiented out).

The Hilbert space of our system seems to be the tensor product of the
Hilbert space of a boson of mass $m^2=\hat e^2{(a-r^2)\over a(a+1-r^2)}$
and of a massless Thirring model; nevertheless the situation is more
intriguing due to the presence of the operator $\psi(x)$ that interpolates
between two extreme situations. We recall that for $x^2\sim 0$, its
behaviour is characterized by the anomalous dimension \eqrefp{eq229} while
the infrared limit is described by the Thirring theory.

We have two critical points corresponding to conformal field
theories in the short and long distance limits: the non critical theory has
both massive and massless degrees of freedom.

The emerging theory, in the large $x$
limit, is not chiral: chirality is in fact screened by the interaction, as the
electric charge is. The short--distance behaviour, on the contrary,
strongly depends on chirality  (as one can see from propagators).
In our case we do not know what
the ultraviolet theory is, if any.
One would
be tempted to think that the ultraviolet theory bears some relations to the
axial--vector generalization of the Thirring model \upcite{Pie92}:
an easy computation
of the critical exponents shows that this is not the case.
We leave this problem open to future investigations.
\vskip -1.0 truecm
\autosez{rem}\underbar{Concluding remarks}

\par
In conclusion we have thoroughly studied a vector--axial vector theory in
two dimensions characterized by a parameter which interpolates between pure
vector and chiral Schwinger models.
The theory has been completely solved by means of non perturbative
techniques, both in a functional approach and in a canonical operatorial
framework.

The main results are the presence of two windows in the space of parameters
in which acceptable  solutions can be obtained. The first window is
characterized by a massive and a massless free bosonic excitations and by
fermions which are endowed with asymptotic states, which feel however a
long range interaction, but in the chiral case.
\noindent
The second window has a massive free boson and a massless ghost; fermions
are confined as their correlators grow with distance. Nevertheless a
Hilbert space of states can be consistently singled out.

The most attractive feature is present in the first window: in this
situation fermionic correlators scale at short and long distances with
different critical exponents.
The infrared limit fully corresponds to a massless Thirring  model times a
free massless bosonic sector. Field, charges and Hilbert space of states do
indeed coincide.
The ultraviolet limit leads to a conformal invariant theory with a larger
number of components (in agreement with Zamolodchikov's theorem
\upcite{Zam86}), whose
Lagrangian formulation, if any, is so far unknown.
These aspects of our model and, more generally, its relation to conformal
invariant theories will be deferred to forthcoming work.
\vskip 1.0 truecm
We are  grateful to A. Johansen for a very stimulating discussion
concerning the scaling properties of our solutions and to F. Strocchi
for useful remarks.

\vfill\eject
\semiautosez{A}\underbar{Appendix A}

\par
In this appendix we show how to derive the left propagator \eqrefp{eq224}
in the path--integral formalism; all the other Green functions can be
obtained in the same way.
\noindent
The first step is to integrate the fermions in \eqrefp{eq24} to give
\eqrefp{eq26} (we put $J_\mu=0$). The change of variables
\eqrefp{eq223} decouples the spinors from $A_\mu$ but has a non trivial
Jacobian ${\cal J}[A_\mu]$
$$
{\cal J}[A_\mu]=\exp\int d^2x{e^2\over
\pi}A_\mu\left[\left(1+a\right)g^{\mu\nu}-\left(1+r^2\right)
{\partial^\mu\partial^\nu\over
\quadratello}-r\epsilon^{\alpha\mu}{\partial_\alpha\partial^\nu\over
\quadratello}\right]A_\nu.
\autoeqno{eqA1}
$$

The fermionic Action is now
$$
\eqalign{
\int d^2x&\left[i\tilde\chi\parz\chi+\bar\eta\exp
ie\left[\alpha+\gamma^5\beta+r\beta+r\alpha\gamma^5\right]\chi\right.+\cr
&\left.+\bar\chi\exp
ie\left[-\alpha+\gamma^5\beta-r\beta+r\alpha\gamma^5\right]\eta\right],\cr}
\autoeqno{eqA2}
$$
\noindent
where $\chi$ is a free fermion and $\alpha, \beta$ are linked by
\eqrefp{eq222} to $A_\mu$.
\noindent
The diagonalization of \eqrefp{eqA2} gives the propagator $S_(x,y;A_\mu)$:
$$
\eqalign{
S_(x,y;A_\mu)=&S^L_0(x-y)\exp\left(i\int
d^2z\, \xi^L_\mu(z;x,y)A^\mu(z)\right)+ \cr
&+S^R_0(x-y)\exp\left(i\int
d^2z\xi^R_\mu(z;x,y)A^\mu(z)\right),\cr
&\xi^{L,R}_\mu(z;x,y)=e(r\pm
1)(\partial^z_\mu\pm\tilde\partial_\mu^z)\left[D(z-x)-D(z-y)\right],\cr}
\autoeqno{eqA3}
$$
\noindent
where $D(x)$ is the free massless scalar propagator in $d=1+1$ and
$S_0^L$, $S^R_0$ the free left and right fermion propagators.

To obtain the left propagator \eqrefp{eq224} we derive
with respect to $\bar\eta_L$ and $\eta_L$
(the left component of the sources \eqrefp{eq25}) and get
$$
S^L(x,y)=S^L_0(x-y)\int{\cal D}  A_\mu{\cal J}[A_\mu]\exp i\int d^2z\left[
-{1\over 4}F_{\mu\nu} F^{\mu\nu}(z)+\xi^L_\mu(z;x,y)A^\mu(z)\right].
\autoeqno{eqA4}
$$

Using the explicit form of ${\cal J}[A_\mu]$ (\eqrefp{eqA1}), we can write the
path--integral over $A_\mu$ as
$$
\int{\cal D} A_\mu \exp\left(i\int d^2 z\left[\xi^L_\mu A^\mu
+{1\over 2}A_\mu K^{\mu\nu}
A_\nu\right]\right),
\autoeqno{eqA5}
$$
\noindent
$K^{\mu\nu}$ being defined in \eqrefp{eq29}.
The Gaussian integration is trivial and gives
$$
S_L(x,y)=S^L_0(x-y)\exp\left(-{1\over 2}\int d^2 zd^2
w\xi^L_\mu(z;x,y)\{K^{-1}\}^{\mu\nu}(z,w)\xi^L_\nu(w;x,y)\right).
\autoeqno{eqA6}
$$

The explicit computation of the exponential factor gives the
renormalization constant $Z_L$ and the interaction contribution in
\eqrefp{eq224}.

\semiautosez{B}\underbar{Appendix B}

\par
We want to investigate in the space of parameters $a$ and $r$, the
limiting
situation $a=r^2$. The Poisson bracket \eqrefp{eq37} vanishes; hence the
request $\Omega_2=0$ implies a third constraint
$$
\Omega_3\equiv -r\Pi_1=0.
\autoeqno{eqB1}
$$

We note that for $r=0$ we have no other constraint in addition to
$\Omega_1 = 0$ and $\Omega_2 = 0$; they are first class and therefore
the theory is gauge invariant.

Obviously $a=r=0$ corresponds to the vector Schwinger model.
Taking $r\not=0$, from
$\dot\Omega_3=0$  we get
$$
\Omega_4\equiv r\left[\hat
e\left(1+r^2\right)A_1-r\partial_1\phi+\Pi_\phi+\hat e rA_0\right]=0.
\autoeqno{eqB2}
$$
Now, since
$$
\left\{\Omega_4(x^1), \Omega_1(y^1)\right\}=\hat e r^2\delta(x^1-y^1),
\autoeqno{eqB3}
$$
we have no further constraints. We end up with a system of four second--class
constraints. Introducing Dirac brackets, we get the non--vanishing
relations
$$
\eqalign{
\left\{A_0(x^1),A_1(y^1)\right\}_D={1+r^2\over r^2\hat
e^2}\partial_x^{1}\delta(x^1-y^1),\qquad&
\left\{A_1(x^1),\phi(y^1)\right\}_D={1\over \hat e}\delta(x^1-y^1),\cr
\left\{A_0(x^1),\phi(y^1)\right\}_D=-{r\over \hat
e}\delta(x^1-y^1),\qquad &\left\{A_1(x^1),\Pi_\phi(y^1)\right\}_D={1\over r\hat
e}\partial_{x^{1}}\delta(x^1-y^1),\cr
\left\{A_0(x^1),\Pi_\phi(y^1)\right\}_D=-{1\over \hat e
r^2}\partial_{x^{1}}\delta(x^1-y^1),\qquad
&\left\{A_1(x^1),A_1(y^1)\right\}_D=-{2\over
r\hat e^2}\partial_{x^{1}}\delta(x^1-y^1),\cr}
$$
$$
\left\{\phi(x^1),\Pi_\phi(y^1)\right\}_D=\delta(x^1-y^1).
\autoeqno{eqB4}
$$
The variables
$\phi$ and $\Pi_\phi$ have a canonical structure and we can express all the
other variables through the constraints to get the Hamiltonian $H_{red}$
$$
H_{red}=\int dx^1\left\{{r^2\over 2}\Pi_\phi^2+{1\over
2r^2}(\partial_1\phi)^2\right\}.
$$

The Heisenberg equations
$$
\eqalign{
&\partial_0\phi=r^2\Pi_\phi,\cr
&\partial_0\Pi_\phi={1\over r^2}\partial_1^2\phi\cr}
\autoeqno{eqB5}
$$
\noindent
are equivalent to
$$
\quadratello\,\phi=0.
\autoeqno{eqB6}
$$
\noindent
The commutation relations are
$$
\eqalign{
&\left[\phi(x),\dot\phi(y)\right]_{E.T.}=ir^2\delta(x^1-y^1),\cr
&\left[\phi(x),\phi(y)\right]_{E.T.}=0. \cr}
\autoeqno{eqB7}
$$

The vector potential is
$$
A_\mu={1\over \hat er}(\partial_\mu-{1\over r}\tilde\partial_\mu)\phi,
\autoeqno{eqB8}
$$
\noindent
giving
$$
F_{\mu\nu}=0,
\autoeqno{eqB9}
$$
\noindent
which is consistent with $\Omega_3=0$.
The only degree of freedom is a massless
scalar excitation.
\noindent
We can construct the fermionic operator solving the Dirac equation in the
same way as in sect.$4$; this time the current coupled to $F^{\mu\nu}$
is zero
and the fermionic sector again describes a Thirring model. The absence of a
massive component in the spectrum forces scale invariance for any $x^2$ :
our model becomes totally equivalent to a massless Thirring model.
\vfill\eject

\biblitem{Thi58} W. Thirring, Ann. of Phys.\underbar{3} (1958) 91.

\biblitem{Sch62} J. Schwinger, Phys. Rev. \underbar{128} (1962) 2425.

\biblitem{Low71} J. H. Lowenstein and J. A. Swieca, Ann. of
Phys.\underbar{68} (1971) 172.

\biblitem{Jac85} R. Jackiw and R. Rajaraman, Phys. Rev. Lett. \underbar{54}
(1985) 1219.

\biblitem{Abd91} E. Abdalla, M. C. B. Abdalla and K. D.Rothe, 2 Dimensional
quantum field theory. World Scientific Singapore 1991.

\biblitem{Hal86} I. G. Halliday, E. Rabinovici, A. Schwirnmer and M.
Chanowitz, Nucl. Phys. \underbar{B268} (1986) 413.

\biblitem{Miy88} S. Miyake and K. Shizuya, Phys. Rev. \underbar{D37} (1988)
2282.

\biblitem{Gam84} R. E. Gamboa Saravi, M. A. Muschietti, F. A. Shaposnik and
J. E. Solomin, Ann. of Phys. \underbar{157} (1984) 360.

\biblitem{Zam86} A. B. Zamolodchikov, JETP Lett. \underbar{43} (1986) 730
and S. J. Nucl. Phys. \underbar{46} (1987) 1090.

\biblitem{Dir50} P. A. M. Dirac, Can. J. Math. \underbar{2} (1950) 129.

\biblitem{Col75} S. Coleman, Phys. Rev. \underbar{D11} (1975) 2088. S.
Mandelstam, Phys. Rev. \underbar{D11} (1975) 3026.

\biblitem{Mor90} G. Morchio, D. Pierotti and F. Strocchi, J. Math. Phys.
\underbar{31} (1990) 1467.

\biblitem{Gir86} H. O. Girotti, H. J. Rothe and K. D. Rothe, Phys. Rev.
\underbar{D33} (1986) 514 and \underbar{D34} (1986) 592.

\biblitem{Itz85} C.Itzykson and J. B. Zuber, Quantum Field Theory, Mc Graw
Hill 1985.

\biblitem{Kla67} B. Klaiber in Lectures in Theoretical Physics, Boulder
1967, Gordon and Breach, New York 1968.

\biblitem{Swi75} B. Schroer, J. A. Swieca and A. H. Volkel, Phys. Rev.
\underbar{D11} (1975) 1509.

\biblitem{Pie92} G. Morchio, D. Pierotti and F. Strocchi, J. Math. Phys.
\underbar{33} (1992) 777.

\biblitem{Fre91} D. Z. Freedman, J. I. Latorre and X. Vilasis, Mod. Phys.
Lett.\underbar{a6} (1991) 531.

\noindent
\underbar{References}
\insertbibliografia
\bye